\documentclass[aps,pra,twocolumn,superscriptaddress,showpacs]{revtex4}
\usepackage{pifont}
\usepackage{txfonts}
\usepackage{amssymb}
\usepackage{graphicx}

\begin{document}

\title{Quantum criticality of the Lipkin-Meshkov-Glick Model in terms
of fidelity susceptibility}

\author{Ho-Man Kwok}%
\affiliation{Department of Physics and Institute of Theoretical Physics, The
Chinese University of Hong Kong, Shatin, Hong Kong, China}

\author{Wen-Qiang Ning}%
\affiliation{Department of Physics and Institute of Theoretical Physics, The
Chinese University of Hong Kong, Shatin, Hong Kong, China}
\affiliation{Department of Physics, Fudan University, Shanghai 200433, China}

\author{Shi-Jian Gu}
 \email{sjgu@phy.cuhk.edu.hk}
\affiliation{Department of Physics and Institute of Theoretical
Physics, The Chinese University of Hong Kong, Shatin, Hong Kong,
China}

\author{Hai-Qing Lin}
\affiliation{Department of Physics and Institute of Theoretical
Physics, The Chinese University of Hong Kong, Shatin, Hong Kong,
China} \affiliation{Department of Physics, Fudan University,
Shanghai 200433, China}

\date{\today}

\begin{abstract}
We study the critical properties of the Lipkin-Meshkov-Glick Model in terms of
the fidelity susceptibility. By using the Holstein-Primakoff transformation, we
obtain explicitly the critical exponent of the fidelity susceptibility around
the second-order quantum phase transition point. Our results provide a rare
analytical case for the fidelity susceptibility in describing the universality
class in quantum critical behavior. The different critical exponents in two
phases are non-trivial results, indicating the fidelity susceptibility is not
always extensive.
\end{abstract}





\maketitle

\section{Introduction}

The Lipkin-Meshkov-Glick (LMG) model \cite{LMG} was introduced in nuclear
physics. It describes a cluster of mutually interacting spins in a transverse
magnetic field. In condensed matter physics, this model is associated with a
system of infinite coordination number. In earlier time, scaling behaviors of
critical observables have been studied by mean field analysis \cite{infcoor},
while recently the finite-size scaling of this model was studied by the $1/N$
expansion in the Holstein-Primakoff single boson representation \cite{HP} and
by the continuous unitary transformations (CUT) \cite{CUT1,CUT2,JVidalStat}.
Meanwhile, a rich structure of four different regions is revealed in the
parameter space through a careful scrutiny on the spectrum
\cite{JVidalSpectra}. Besides, the quantum criticality has been investigated by
studying its entanglement properties \cite{QPT1st,QPT2nd,EntE1,EntE2,nilesen}.
Both the first- and second-order quantum phase transitions (QPTs)
\cite{sachdev} have been revealed, in the antiferromagnetic and the
ferromagnetic cases respectively \cite{QPT1st,QPT2nd}.

Regarding the QPT itself, the ground state of a system would
undergo a significant structural change at certain critical point.
This primary observation suggests a new description of QPTs in
terms of fidelity
\cite{HTQuan06,PZanardi06,Buonsante07,PZanardi0606130,PZanardi07,WLYou07,HQZhou07,LCVenuti07,SJGu07,SChen07,WQNing07,MFYang07,NPaunkovic07},
a concept introduced in quantum information theory \cite{nilesen}.
Mathematically it is the overlap between two ground states in
which their driving parameters deviate by a small amount.
However, the fidelity depends computationally on an arbitrarily
small yet finite change of the driving parameter. For this,
Zanardi \emph{et. al.} introduced the Riemannian metric tensor
\cite{PZanardi07}, while You \emph{et. al.} suggested the
fidelity susceptibility \cite{WLYou07}, both focus on the leading
term of the fidelity, in order to explain singularities in QPTs.
In addition, scaling analysis of these quantities has been
informative: it helps understanding their divergence and the
criticality of the system \cite{LCVenuti07}, and it also reveals
the intrinsic relation between the critical exponent of some
physical quantities and that of the fidelity susceptibility
\cite{SJGu07}.

In this paper, we explicitly compute the ground-state fidelity susceptibility
and its critical exponent of the LMG model. Numerical analysis
is also performed to check with our analytic calculations. We show that, the
$1/N$ expansion in the Holstein-Primakoff transformation is sufficient to determine
the critical exponent of the fidelity susceptibility $\chi_{_F}$. In addition, we
revealed two distinct critical exponents in two phases which is not a general
feature. Therefore, our findings not only suggest another route on understanding
the quantum criticality of the LMG model, but also show the fidelity susceptibility
is not always extensive in describing the universality class of a quantum many-body
system.

This paper consists of five sections. In Sec. \ref{sec:Ham}, we review the
Hamiltonian, symmetry, and conserved quantities of the LMG model. In Sec.
\ref{sec:criticalfs}, we diagonalize the model Hamiltonian and compute the
fidelity susceptibility in the anisotropic model. In Sec. \ref{sec:scalingana},
we perform finite size scaling analysis and discuss the
scaling relation between different exponents. Finally, we give a brief summary
in Sec. \ref{sec:sum}.

\section{The model Hamiltonian}
\label{sec:Ham}

The Hamiltonian of the LMG model reads
\begin{eqnarray}
 H &=&  - \frac{\lambda }{N}\sum\limits_{i < j} {\left( {\sigma _x^i \sigma _x^j
 + \gamma \sigma _y^i \sigma _y^j } \right)}  - h\sum\limits_i {\sigma _z^i
 },
 \\
 &=&  - \frac{{2\lambda }}{N}\left( {S_x^2  + \gamma S_y^2 } \right) - 2hS_z
 + \frac{\lambda }{2}\left( {1 + \gamma } \right),
 \\
 &=&  - \frac{\lambda }{N}\left( {1 + \gamma } \right)\left( {\textbf{S}^2  - S_z^2  - N/2} \right) - 2hS_z
 \nonumber \\ && - \frac{\lambda }{{2N}}\left( {1 - \gamma } \right)\left( {S_ + ^2  + S_ - ^2 } \right)
 , \label{eq:HLMG}
\end{eqnarray}
where $\sigma _{\kappa}\, (\kappa=x,y,z)$ are the Pauli matrices,
$S_\kappa = \sum _i \sigma _{\kappa}^i/2$, and $S_\pm  = S _x \pm
iS _y$. The prefactor $1/N$ is necessary to ensure finite energy
per spin in the thermodynamic limit. It is understood that the
total spin and the parity are the conserved quantities, i.e.,
\begin{eqnarray}
\left[ {H,S^2 } \right] = \left[ {H,\prod\limits_i {\sigma _z^i } }
\right] = 0  .
\end{eqnarray}
In addition, in the isotropic case $\gamma = 1$, one has $\left[ {H,S_z }
\right] = 0$ and simultaneous eigenstates can be found. In the main context,
the following parameter space is considered: $\lambda = 1, |\gamma| < 1, h
\geq 0$. We take $h \geq 0$ as the spectrum is invariant under the transformation
$h\leftrightarrow-h$. In addition, as a common practice we only consider the
maximum spin sector $S = N/2$ which contains the lowest energy state.

\section{Critical behavior of the fidelity susceptibility}
\label{sec:criticalfs}

We briefly review of the concept of the fidelity susceptibility
here. Suppose there is a Hamiltonian of a general form as
\begin{eqnarray}
H = H_0(\gamma) + f(h) H_I, \label{eq:generalHam}
\end{eqnarray}
for $H_I$ is defined as the driving term of the system, which
simply does not commute with $H_0$. The function $f(h)$ coupled
to $H_I$ is often considered as the linear external field
$f(h)=h$. Then the fidelity susceptibility is defined as
\cite{PZanardi07,WLYou07}
\begin{eqnarray}
\chi_{_F} =\left[\frac{d f(h)}{dh}\right]^2
\sum\limits_{n \ne 0} {\frac{{\left| {\left\langle {n} \right|H_I
\left| {0}  \right\rangle } \right|^2 }}{{\left[ {E_n  - E_0 }
\right]^2 }}}. \label{eq:definitionfs}
\end{eqnarray}
where $E_n$ and $|n\rangle$ stand for the $n^{\rm th}$
eigenenergies and eigenstates of the (whole) Hamiltonian
respectively.

The fidelity susceptibility is well-defined for a non-degenerate
ground state of the continuous variable $h$, but it is not suitable
to deal with states with good quantum numbers. The LMG model undergoes
ground state level crossing when $\gamma = 1$, the ground states are
assigned the magnetization as the quantum numbers.

We put our focus on the fidelity susceptibility for an arbitrary isotropy
$|\gamma| < 1$. One resolution is to use the Bethe-Ansatz solution
\cite{FPan99,JLinks03}, which is rather complicated. So we adopt the
$1/N$ expansion method which was used extensively by Dusuel and
Vidal \cite{CUT1,CUT2}, that corresponds to the large $N$ limit.

The $1/N$ expansion method is done under the Holstein-Primakoff
boson representation \cite{HP} framework. In low energy spectrum
the spin operators in the $S = N/2$ subspace are mapped into
boson operators:
\begin{eqnarray}
 S_z &=& S - a^{\dagger}a, \nonumber
 \\
 S_+ &=& (2S-a^{\dagger}a)^{1/2}a= N^{1/2}(1-a^{\dagger}a/N)^{1/2}a=S_-^{\dagger},
\end{eqnarray}
where $a$($a^{\dagger}$) is the standard bosonic annihilation
(creation) operator satisfying $[a,a^{\dagger}]=1$. The above
transformation is valid when $h\geq 1$, but when $0<h<1$ it can
also be used through semi-classical treatment \cite{CUT1,CUT2}.
This representation is also known as the spin-wave theory. It is
well adapted to the computation of the low-energy physics when
$\langle a^{\dagger}a\rangle/N \ll 1$. After inserting these
expressions of the spin operators in Eq. (\ref{eq:HLMG}),
one can approximate the square roots as one and express the
result in normal ordered form with respect to the boson vacuum
state. Keeping terms of order $(1/N)^{-1}$, $(1/N)^{-1/2}$ and
$(1/N)^{0}$ for $h\geq 1$ (in which the approximation is
justified), the Hamiltonian becomes
\begin{eqnarray}
H = -hN  + (2h-1+\gamma)a^{\dagger}a -\frac{1-\gamma}{2} \left(
a^{\dagger}{^2}+a^2 \right).
\end{eqnarray}
The above Hamiltonian can be diagonalized by a standard Bogoliubov
transformation
\begin{eqnarray}
a^{\dagger}&=&\cosh(\Theta/2)b^{\dagger}+\sinh(\Theta/2)b,\\
a &=& \sinh(\Theta/2)b^{\dagger}+\cosh(\Theta/2)b,
\end{eqnarray}
where $b (b^\dagger)$ is the quasi-bosonic annihilation (creation)
operator, and
\begin{eqnarray}
\tanh[\Theta(h\geq 1)]=\frac{1-\gamma}{2h-1+\gamma},
\end{eqnarray}
then the Hamiltonian is diagonalized as
\begin{eqnarray}
H = -h(N+1)+
2\sqrt{(h-1)(h-\gamma)}\left(b^{\dagger}b+\frac{1}{2}\right).
\end{eqnarray}
Thus the low-energy spectrum of the model is mapped to the
spectrum of a simple harmonic oscillator. The eigenstates are just
$\{|n\rangle\}$, where $b^{\dagger}b|n\rangle=n|n\rangle$. We
consider the driving Hamiltonian $H_I$ responsible for the QPT,
\begin{eqnarray}
H_I = -\sum\limits_i{\sigma_z^i}&=&-2S_z.
\end{eqnarray}
By transforming them into combinations of $b$ and $b^\dagger$
operators, the fidelity susceptibility is calculated as
\begin{eqnarray}
\chi_{_F} = \frac{(1-\gamma)^2}{32(h-1)^2(h-\gamma)^2}. \label{eq:FShg1}
\end{eqnarray}

The derivation above is only valid for $h\geq1$, for $0<h<1$ the
calculation is actually similar to the above case of $h\geq1$,
provided that one first rotates the $z$ axis to bring it along
the classical spin direction. We do not show it explicitly here,
but interested readers are recommended to refer to Ref.
\cite{CUT1,CUT2}. We simply quote the main result, after all the
procedures the Hamiltonian becomes:
\begin{eqnarray}
H = -\frac{(1+h^2)}{2}N-\frac{1-\gamma}{2} +
2\sqrt{(1-h^2)(1-\gamma)}\left(b^\dagger b + \frac{1}{2} \right).
\end{eqnarray}
The driving Hamiltonians also takes a different form:
\begin{eqnarray}
-\sum\limits_i{\sigma_z^i} &=& -2S_z \nonumber \\
&=& -2\left(-\sqrt{1-h^2}\widetilde{S_x}+h\widetilde{S_z}\right),
\end{eqnarray}
for the HP transformation is done on the $\widetilde{S}$ operators.
The fidelity susceptibilities are then obtained accordingly:
\begin{eqnarray}
\chi_{_F} &=&
\frac{N}{4\sqrt{(1-h^2)(1-\gamma)}}+\frac{h^2(h^2-\gamma)^2}{32(1-\gamma)^2(1-h^2)^2}.
\label{eq:FShl1}
\end{eqnarray}

Thus we obtained $\chi_{_F}$ of the anisotropic LMG model in large $N$ limit.
We first see the effect of isotropy to the fidelity susceptibility. It
dominates when $h<1$, but fades out for large $h$. Especially in the isotropic
limit, when $\gamma \to 1$, $\chi_{_F}$ diverges when $h<1$, but tends to zero when $h>1$.
This is the effect of the level-crossing points in the thermodynamic
limit. They together form a region of criticality, and the system undergoes
continuous level crossing. The fidelity susceptibility responds drastically while moving
along $h$. But when $h>1$, there are no further critical points,
$\chi_{_F}$ naturally measures zero when moving along $h$ because we have
$[H_0,H_I]=0$.

\begin{figure}
\includegraphics[width=7cm]{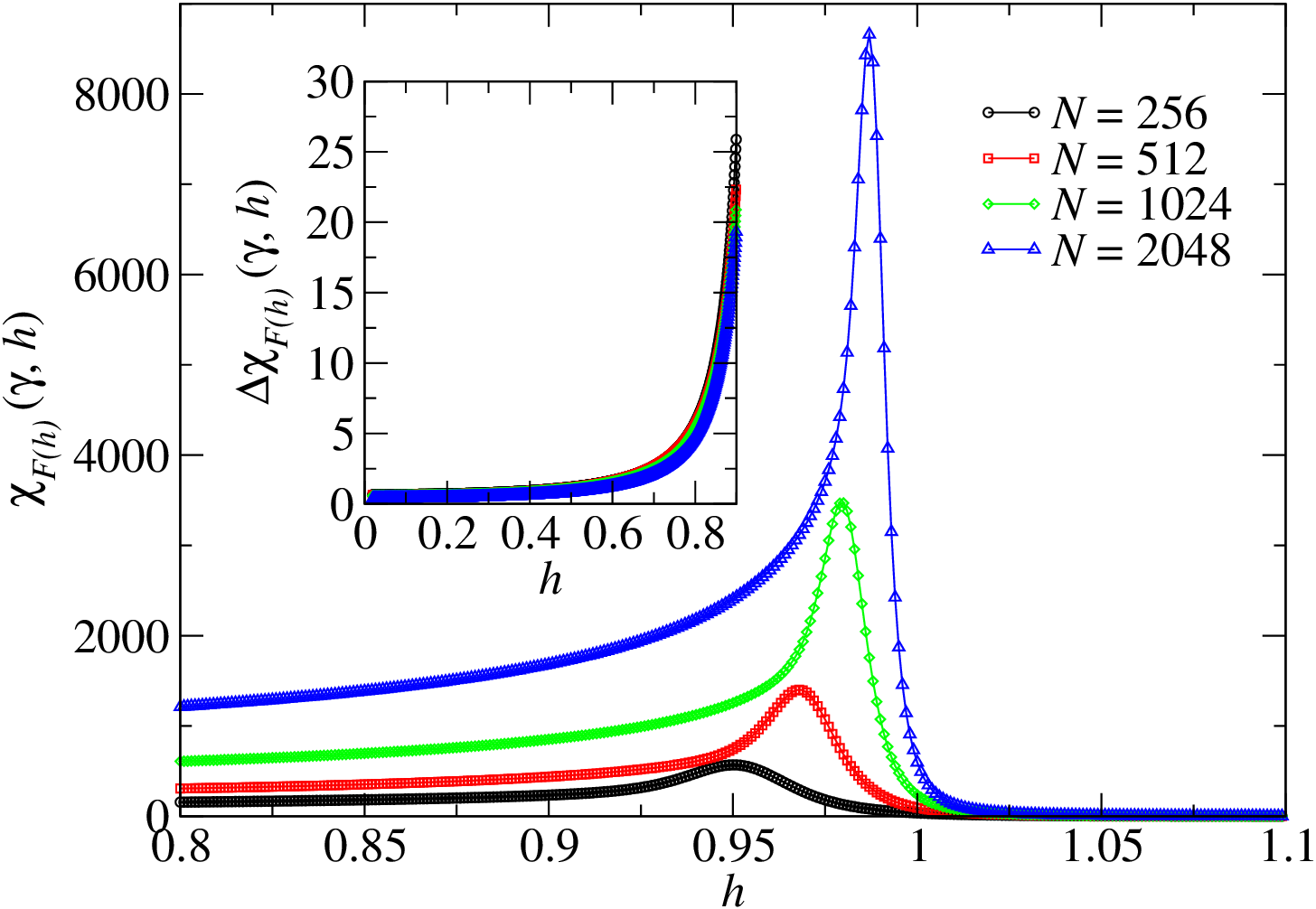}
\caption{(color online) The fidelity susceptibility in response to $h$ as a
function of $h$ at $\gamma = 0.5$. The inset denotes the difference between the
fidelity susceptibility and the extensive term in Eq. (\ref{eq:FShl1}).}
\label{figure_fs_h05}
\end{figure}

An interesting observation is $\chi_{_F}$ behaves extensively when $h<1$
even in the large $N$ limit. When discarding the extensive part of Eq.
(\ref{eq:FShl1}), we arrive a zero point at $h = \sqrt{\gamma}$, which does
not fit with numerical analysis [Fig. \ref{figure_fs_h05}]. This discrepancy
may be eliminated by adopting other transformations of the driving Hamiltonian.
Particularly, the flow of operators in the LMG model haven been studied by the
continuous unitary transformation (CUT) method \cite{CUT1,CUT2}.
However, such discrepancy would not hinder us from getting the correct critical exponent
of the fidelity susceptibility.

Let us emphasize the intensive property of the fidelity susceptibility, which
measures the average response to some driving Hamiltonians. Its divergence
should correspond to a critical point of a second-order QPT rather than to the
increasing system size. In order to predict the critical exponent correctly,
we should average the fidelity susceptibility whenever necessary. To the leading
order, Eq. (\ref{eq:FShl1}) becomes
\begin{eqnarray}
\frac{\chi_{_F}}{N} = \frac{1}{4\sqrt{(1-h^2)(1-\gamma)}}.
\end{eqnarray}
Then it comes to a key result of our paper: $\chi_{_F}$ bears
different critical exponents across the critical point. It diverges as
$(1-h)^\frac{1}{2}$ when $h<1$, $(h-1)^2$ when $h>1$. It is unlike
the Ising model in a transverse field \cite{PZanardi06} nor the one-dimensional
asymmetric Hubbard model \cite{SJGu07}, where the critical exponent is a
single number over the phases.

\begin{figure*}
\includegraphics[width=5.5cm]{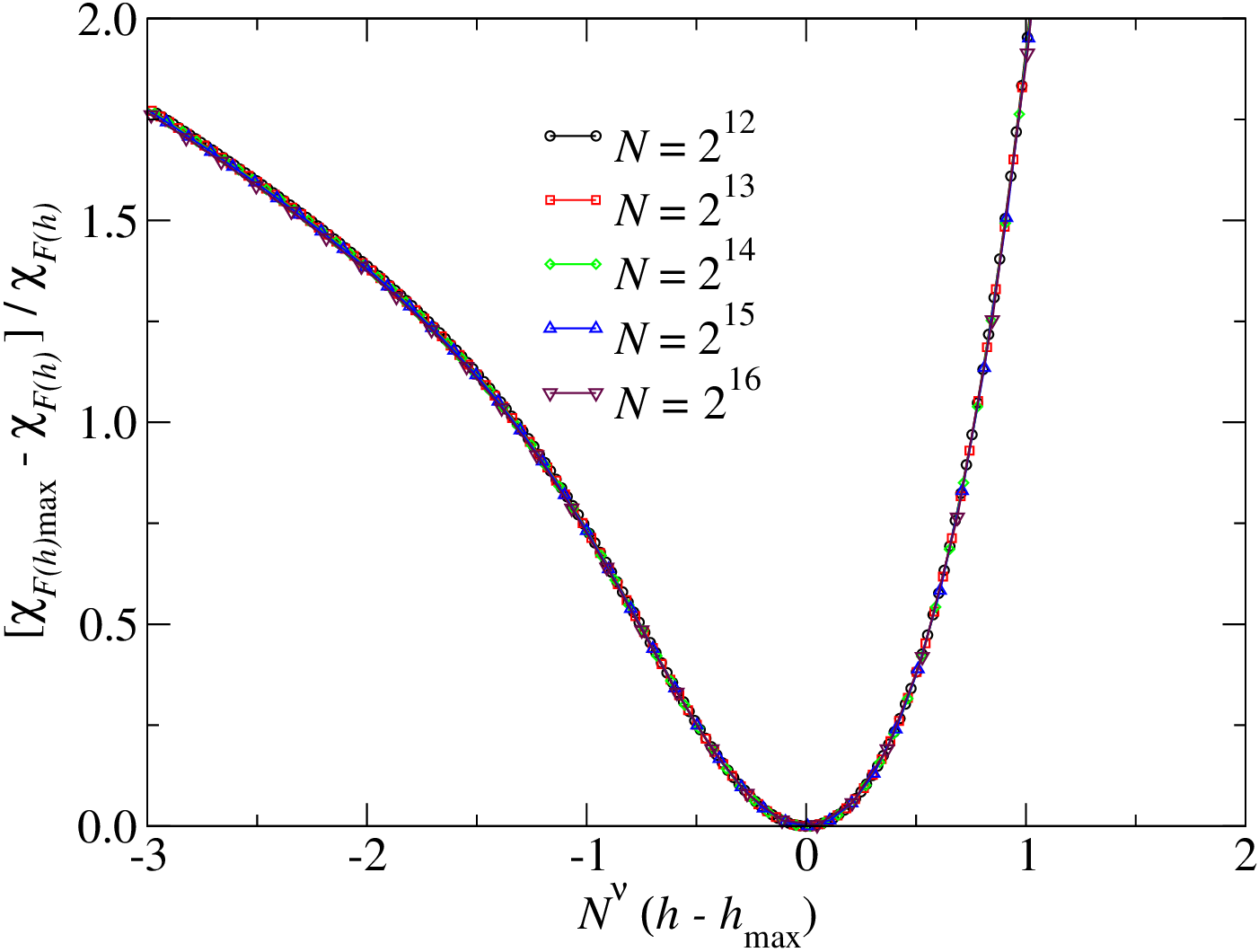}\hspace{3mm}
\includegraphics[width=5.5cm]{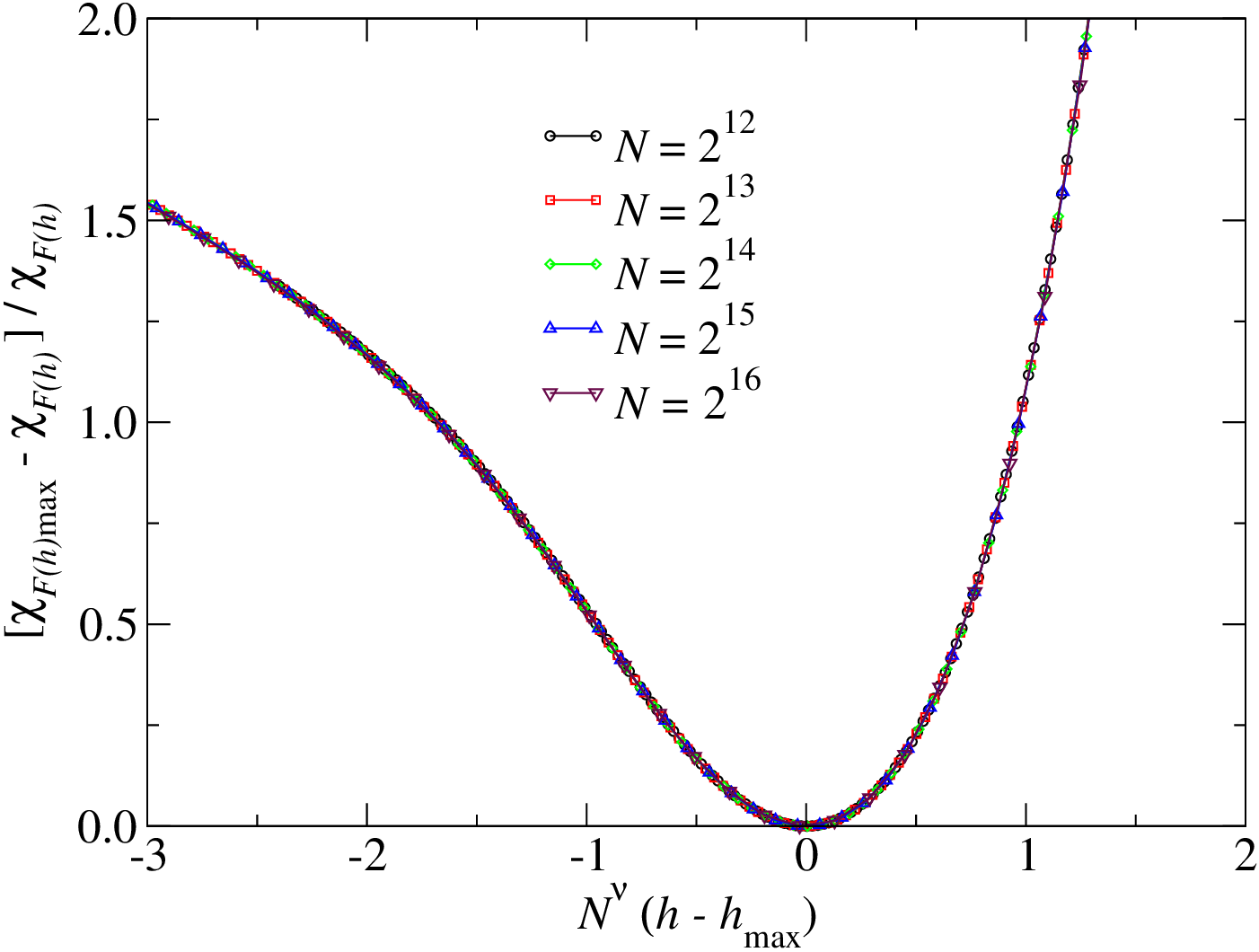}\hspace{3mm}
\includegraphics[width=5.5cm]{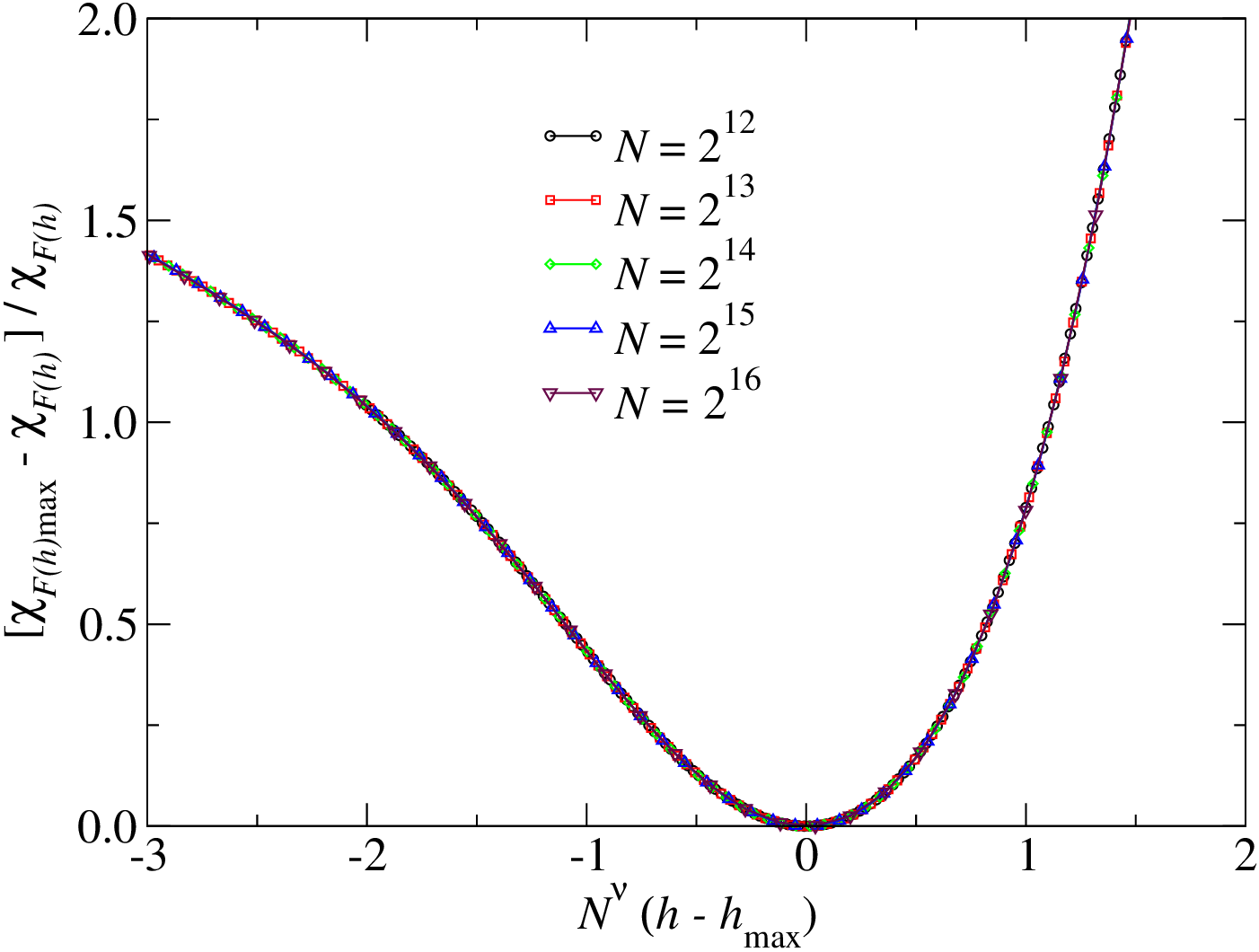}
\caption{(color online) The finite size scaling analysis is
performed for the case of power-law divergence at $\gamma = 0.5$
(LEFT), $\gamma = 0$ (MIDDLE) and $\gamma = -0.5$ (RIGHT) for system
sizes $N=2^{n} (n=12, 13, 14, 15, 16)$. The fidelity
susceptibility is considered as a function of system size and
driving parameter is a function of $N^\nu (h-h_{\rm max})$ only,
with the correlation length critical exponent $\nu\simeq 0.665$.} \label{figure_3}
\end{figure*}

\begin{figure}
\includegraphics[width=7cm]{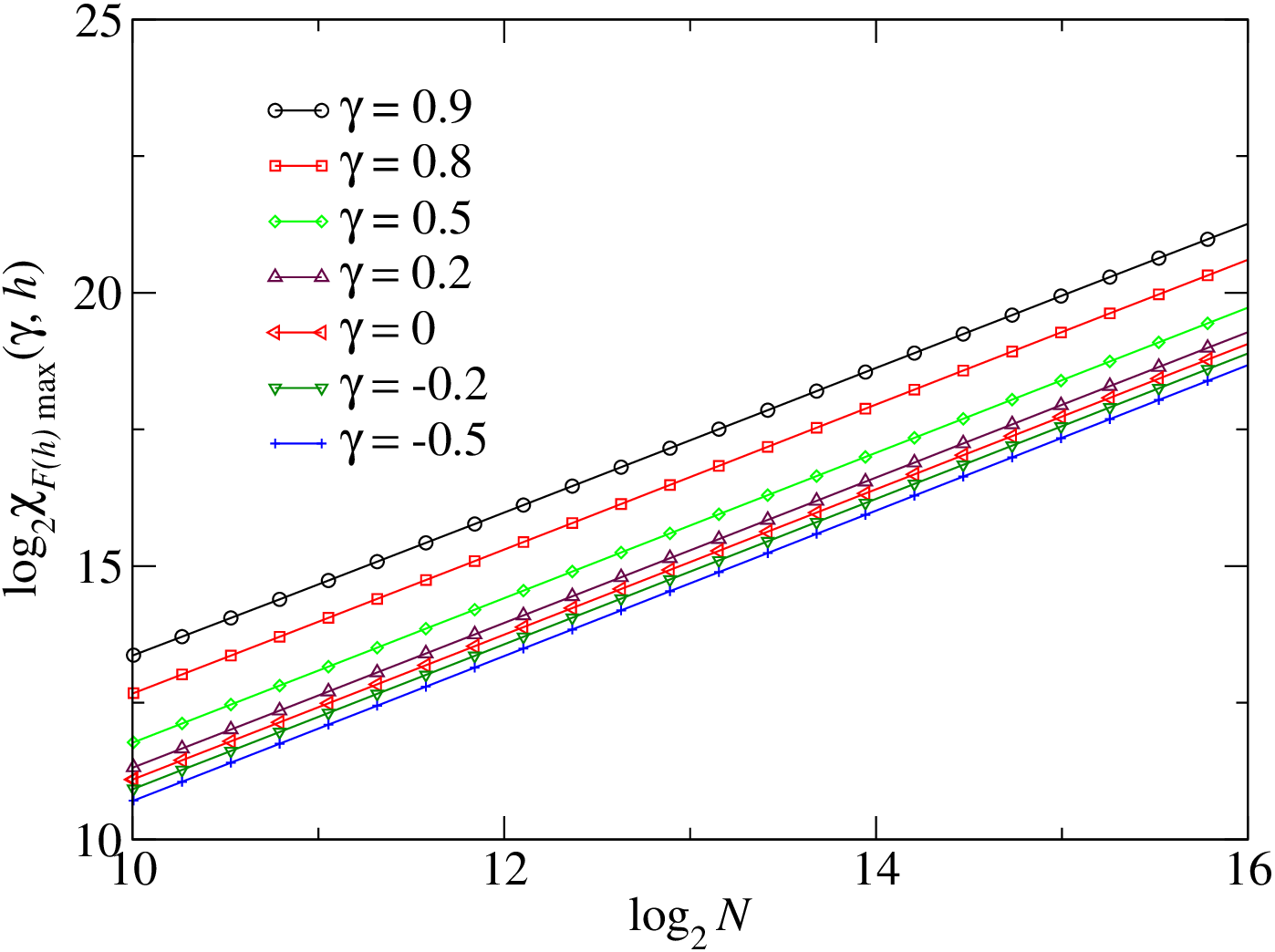}
\caption{(color online) The finite size scaling is performed for
the maximum of the fidelity susceptibility.} \label{figure_4}
\end{figure}

\section{Finite size scaling analysis}
\label{sec:scalingana}

To illustrate the scaling behavior of the fidelity
susceptibility, we perform the exact diagonalization (ED) to
solve the spectrum of $H$ and then calculate the corresponding
fidelity susceptibility numerically.

Let us recall the fidelity susceptibility scaling analysis performed in
the asymmetric Hubbard model \cite{SJGu07}. According to the scaling ansatz
\cite{MAContinentinob} and the obvious power-law divergence observed in Fig.
\ref{figure_fs_h05}, the rescaled fidelity susceptibility
around its maximum point at $h_{\rm max}$ is a simple function of a scaling variable,
i.e.
\begin{eqnarray}
\frac{\chi_{_F{\rm max}} - \chi_{_F}}{\chi_{_F}} = f[N^\nu (h-h_{\rm max})],
\label{eq:scalingansatz}
\end{eqnarray}
where $f(x)$ is the scaling function and $\nu$ is the correlation
length critical exponent. This function is universal and does not
depend on the system size, as shown in Fig. \ref{figure_3} for
cases of $\gamma = 0.5, 0$ and $\gamma = -0.5$. Remarkably, the critical
exponent $\nu$ for three cases are very close. This observation
strongly implies that $\nu$ is a universal constant and does not
depend on the parameters $\gamma$ and $h$.

\begin{table*}
\caption{Scaling exponent $\mu$ at various $\gamma$, obtained by sampling
 system size in different range.} \label{tab:critcalexp}
\begin{center}
\begin{ruledtabular}
\begin{tabular}{ccccccc}
 $\gamma$& 0.8 & 0.5 & 0.2 & 0 & -0.2 & -0.5 \\
\hline $\mu (N\in[2^8, 2^{16}])$ & $1.3221\pm 0.0006$ & $1.3264\pm 0.0004$ &
$1.3267\pm 0.0004$ & $1.3280\pm 0.0004$
& $1.3283\pm 0.0003$ & $1.3285 \pm 0.0003$ \\
\hline $\mu (N\in[2^{12}, 2^{16}])$ & $1.3250\pm 0.0003$ & $1.3285\pm 0.0004$ &
$1.3295\pm 0.0002$ & $1.3299\pm 0.0002$
& $1.3302\pm 0.0001$ & $1.3304 \pm 0.0001$\\
\end{tabular}
\end{ruledtabular}
\end{center}
\end{table*}

In recent studies of the fidelity susceptibility in critical
phenomena, it was pointed out that the intensive fidelity
susceptibility scales generally like \cite{LCVenuti07,SJGu07}
\begin{eqnarray}
\chi_{_F} \propto \frac{1}{|h-h_c|^\alpha},
\end{eqnarray}
around the critical point. In the last section, we have already obtained
\begin{eqnarray}
\alpha  = \left\{ \begin{array}{ccc}
 2,& &h > 1 \\
 \frac{1}{2}, & & 0 \le h < 1 \\
 \end{array} \right. ,
\end{eqnarray}
which is also a universal constant. Then if the maximum point of
the intensive fidelity susceptibility scales like
\begin{eqnarray}
\chi_{_F{\rm max}} \propto N^\mu,
\end{eqnarray}
the scaling ansatz also implies another important relation, i.e.
\begin{eqnarray}
\alpha=\frac{\mu}{\nu} . \label{eq:scalingequality}
\end{eqnarray}

We try to confirm this equality in numerically. In Fig. \ref{figure_3}, Eq.
(\ref{eq:scalingansatz}) is best fitted with $\nu\simeq 0.665$. The case to
determine $\mu$ is more subtle. It is because Eq. (\ref{eq:scalingansatz})
remains the same form even for averaged $\chi_{_F}$, but the maximum of
$\chi_{_F}$ does not. To resolve this problem, we first determine $\mu$
from the ``bare" $\chi_{_F}$. By using least square fit method, we
evaluated ``bare" $\mu$ for at different $\gamma$. The numerical details are
shown in table \ref{tab:critcalexp}. However, the exponent $\mu$ does not
converge perfectly. We compare the $\mu$ obtained in a range of $[2^{12},
2^{16}]$, and those from the range $[2^{8}, 2^{16}]$. The results converge
better for larger scaling regions. According to the trend of $\mu$ in larger
system sizes, we roughly estimate $\mu=1.33$ with three effective digits [Fig.
\ref{figure_4}].

When $h>1$, $\chi_{_F}$ is observed to be intensive [Fig.
\ref{figure_fs_h05}]. With the estimated $\mu$ and $\nu$, the equality
(\ref{eq:scalingequality}) is satisfied with $\alpha=2$. On the other hand,
when $h<1$, $\chi_{_F}/N$ is the intensive quantity. For $\chi_{_F}
\propto N^\mu$,
\begin{eqnarray}
\frac{\chi_{_F}}{N} \propto N^{(\mu-1)}.
\end{eqnarray}
Thus $\mu \simeq 0.33$, this will give the relation $\alpha =
1/2$. These two values of $\alpha$ are consistent with our analytic
calculation in the last section.

The exponent $\mu$, $\nu$ can also be discussed from the scaling
ansatz at the critical point rather than the maximum point of a
finite system, as shown by Vidal, Dusuel, and Barthel
\cite{CUT2,JVidalStat}. Based on their approach, the critical
exponent $\nu$ takes the value of 1/3, and is independent of magnitude
of $\gamma$. Our results on the maximum simply agree with this
value and can be generalized to other models where the precise critical
point is not known.

Another scaling analysis is to examine how $h_{\rm max}$ tends to
the critical point $h_c=1$. It should scale like
\begin{eqnarray}
h_c - h_{\rm max} \propto N^{\,-\delta},
\end{eqnarray}
in the large $N$ limit. We find $\delta \simeq 0.66$ with two
effective digits for various $\gamma$.

In short, we can confirm that the exponents $\mu, \nu$, and
$\delta$ of the fidelity susceptibility do not depend on the
value of $\gamma$ and $h$. They are universal constants for the LMG
model and are related to the critical exponent of the fidelity susceptibility $\alpha$.

\section{Summary}
\label{sec:sum}

In summary, we computed explicitly the fidelity susceptibility and its
critical exponent of the LMG model at different isotropy.
We confirmed the different critical exponents in two phases numerically by
ED, which is a rather non-trivial result. Several scaling exponents are
also found in consistence with previous studies. Since the fidelity susceptibility is believed to be
able to characterize the universality class of quantum phenomena,
our results therefore provide a rare explicit case for the study of
fidelity susceptibility.

\begin{acknowledgements}

We are very grateful to J. Vidal for many fruitful comments. S. J. G. thanks X.
Wang and J. P. Cao for helpful discussions. This work was partially supported
RGG Grant CUHK 400906, 401504, and MOE B06011.


\end{acknowledgements}

\end{document}